\documentclass[useAMS,usenatbib,epsfig]{mn2e}

\newif\ifAMStwofonts
\AMStwofontstrue
\usepackage{amsmath,amsfonts,epsfig,natbib,longtable}

\def\reff@jnl#1{{\rm#1\/}}
\def\aj{\reff@jnl{AJ}}                  
\def\araa{\reff@jnl{ARA\&A}}            
\def\apj{\reff@jnl{ApJ}}                        
\def\apjl{\reff@jnl{ApJ}}               
\def\apjs{\reff@jnl{ApJS}}              
\def\ao{\reff@jnl{Appl.Optics}}         
\def\apss{\reff@jnl{Ap\&SS}}            
\def\aap{\reff@jnl{A\&A}}               
\def\aapr{\reff@jnl{A\&A~Rev.}}         
\def\aaps{\reff@jnl{A\&AS}}             
\def\azh{\reff@jnl{AZh}}                        
\def\baas{\reff@jnl{BAAS}}              
\def\gca{\reff@jnl{GeCoA}}              
\def\jrasc{\reff@jnl{JRASC}}            
\def\memras{\reff@jnl{MmRAS}}           
\def\mnras{\reff@jnl{MNRAS}}            
\def\pra{\reff@jnl{Phys.Rev.A}}         
\def\prb{\reff@jnl{Phys.Rev.B}}         
\def\prc{\reff@jnl{Phys.Rev.C}}         
\def\prd{\reff@jnl{Phys.Rev.D}}         
\def\prl{\reff@jnl{Phys.Rev.Lett}}      
\def\pasp{\reff@jnl{PASP}}              
\def\pasj{\reff@jnl{PASJ}}              
\def\qjras{\reff@jnl{QJRAS}}            
\def\skytel{\reff@jnl{S\&T}}            
\def\solphys{\reff@jnl{Solar~Phys.}}    
\def\sovast{\reff@jnl{Soviet~Ast.}}     
\def\ssr{\reff@jnl{Space~Sci.Rev.}}     
\def\zap{\reff@jnl{ZAp}}                        
\def\nat{\reff@jnl{Nature}}             

\title[The spatial distribution of galaxies within the CMB cold spot in the CrB-SC]{The spatial
distribution of galaxies within the CMB cold spot in the Corona Borealis supercluster}
\author[Padilla-Torres et al.] {Carmen Pilar Padilla-Torres,$^{1}\thanks{E-mail: cppt@iac.es}$ Carlos
M. Guti\'errez,$^{1}$ Rafael Rebolo,$^{1,2}$ 
\newauthor Ricardo G\'enova-Santos,$^{1}$ and  Jos\'e Alberto Rubi\~no-Martin$^{1}$ 
\\
$^1$ Instituto de Astrofis\'{\i}ca de Canarias, 38200 La Laguna, Tenerife, Canary Islands, Spain \\
$^2$ Consejo Superior de Investigaciones Cient\'{\i}ficas, Spain }  
\date{Accepted Received In original form}
\pagerange{\pageref{firstpage}--\pageref{lastpage}}
\pubyear{2008a}

\begin{document}

\label{firstpage}
\maketitle

\begin{abstract}

We study the spatial distribution and colours of galaxies within the region covered by the
cold spot in the cosmic microwave background (CMB) recently detected by the Very Small Array
(VSA; G\'enova-Santos et al.\ 2005, 2008) towards the Corona Borealis supercluster (CrB-SC).
The spot is in the northern part of a region  with a radius $\sim 1$ degree ($\sim 5$ Mpc at
the redshift of CrB-SC) enclosing the clusters Abell  2056, 2065, 2059 and 2073, and
where the density of galaxies, excluding the contribution from those clusters, is $\sim 2$
times higher  than the mean value in typical intercluster regions of the CrB-SC. Two of such
clusters (Abell  2056 and 2065) are members of the CrB-SC, while the other two are in the
background.  This high density intercluster region is quite inhomogeneus, being the most
remarkable feature  a large concentration of galaxies in a narrow  filament running from
Abell  2065 with a length of $\sim 35$ arcmin ($\sim 3$ Mpc at the redshift of CrB-SC)  in the
SW--NE direction. This intercluster population of galaxies probably results from the interaction
of  clusters Abell  2065 and 2056. The area subtended by the VSA cold spot shows an excess of
faint ($21<r<22$) and red  ($1.1<r-i<1.3$) galaxies as compared with typical values within
the CrB-SC intercluster regions. This overdensity of galaxies shows a radial dependence and
extends out to $\sim 15$ arcmin. This could be  signature of a previously unnoticed cluster 
 in the background. 
\end{abstract}

\section{Introduction}

Observations of the Corona Borealis supercluster (hereafter CrB-SC) with the Very Small
Array (VSA)  extended configuration (G\'enova-Santos et al. 2005, hereafter GS05) at
33~GHz have shown  a  decrement  in the temperature of the  cosmic microwave background (CMB),
 with a minimum temperature of $-230\pm 23~\mu$K  at coordinates RA$=15^{\rm h}22^{\rm
m}11.47^{\rm s}$, Dec.$=+28\degr 54'06.2''$ (J2000), near the center of the  supercluster, in a regionwith no known galaxy clusters. The explanation of such a strong temperature 
decrement based only on a primary Gaussian  CMB anisotropy is extremely unlikely.
 This led us to consider the Sunyaev-Zel'dovich
(SZ)  effect  generated either by a previously unnoticed galaxy cluster or by a
filamentary structure consisting  of warm/hot gas in this supercluster  as a possible
component of this decrement.  The SZ  effect (Sunyaev \& Zeldovich 1972) arises from the
inverse  Compton scattering  of the CMB photons by hot
electrons and produces decrements in the mean  temperature of the CMB at the VSA
frequency. It has been detected in the direction of a number of galaxy  clusters 
(see e.g.  Lancaster et al.'s 2005 for  VSA observations of this effect in galaxy clusters);
however,  large-scale structures in superclusters, with lower temperatures and baryon
overdensities but much larger  length scales, may also generate detectable SZ signals. 

Millimetric observations with the MITO telescope at 143, 214
and 272~GHz  (Battistelli et al.\ 2006)  supported
the presence of an SZ component, probably on top  of a negative primary CMB anisotropy.
Recently we have also  carried out observations with the new superextended
configuration of the VSA (G\'enova-Santos et al.\ 2008,  hereafter G08), completely
independently and at a different angular resolution of 7 arcmin (FWHM), which confirmed 
the presence of this decrement. The gaussianity analyses performed on these data also
bore out the results of GS05.  In that work we also showed that this decrement is also
detected in the WMAP 5-year data, even though with a lower  level of signifiance
because of its coarser angular resolution and higher noise.

The joint restrictions set by the VSA decrement and by the absence of significant X-ray
signal in the  ROSAT {\it All Sky Survey} data (see fig.~9 of GS05) put strong
constraints on the temperature and density of  the gas responsible for such a decrement.
In GS05 we estimated that only $\sim$0.3 galaxy clusters rich enough to produce  such a
deep decrement are expected in the entire surveyed region. Also, this cluster
hypothesis is further disfavoured by the fact that, due to the absence of any X-ray
excess, a distant cluster would be needed, and this is in disagreement with the large
angular extension of the decrement. On the other hand, as we showed in GS05, a 
filamentary structure of diffuse gas pointing towards us with a temperature 0.6-0.8~keV,
baryon overdensity $\sim$400 and length $\sim$40~Mpc, could account for  50\% of the
total detected decrement whilst remaining undetected in the ROSAT data. This filament
would contain a gas mass $\sim 5\times 10^{14}$ M$\sun$.
Such structures  could provide a  location for a significant fraction of the baryon content 
in the local Universe.

 In fact, the baryon density at $z=0$, derived from the total budget over the
well-observed components (Fukugita et al.\ 1998) is a factor of $\approx$2 lower than
that at high redshift, inferred through independent methods, such as the Ly$\alpha$
forest (e.g. Rauch et al.\ 1997), the primordial CMB fluctuations (e.g. Rebolo et al.\
2004; Spergel et al.\ 2005) or the primordial abundance of deuterium (Tytler et al.\
1996).  The results of hydrodynamical simulations  (Cen \& Ostriker 1999; Dav\'e et al.
1999, 2001) indicate that a significant fraction of these ``missing baryons'' may be
located in a phase of warm/hot gas distributed in diffuse structures on supercluster
scales, at temperatures  $10^{5}\leq T\leq 10^{7}$ K. Basilakos et al.\ (2006) have
studied the morphology of superclusters by means of  hydrodynamical and $N$-body
simulations in a $\Lambda$-CDM model. They found that filamentary structures are very
prominent in clusters and superclusters dominated by dark matter. The gaseous component
is, however, predominantly spherical. 

If the decrement detected by the VSA is caused by an SZ effect arising from a filamentary
structure of warm/hot gas along the line of sight, there should be galaxies towards
this direction which may be tracing the distribution of this gas. Therefore, an optical
study of the galaxy population in this region is useful for determining whether the 
hypotethical SZ effect airses from such a structure or from a farther unknown galaxy
cluster. In this paper we analyse the density,
spatial distribution and photometric properties  of such galaxies using the Sloan
Digital Sky Survey (SDSS). The paper is organized as follows: Section~2 describes the
main  properties of the CrB-SC from an optical perspective; Section 3 presents the
basic properties of the microwave and optical data used in our study; Section~4
presents the analysis of such data, focusing on  the properties of the spatial
distribution  of galaxies  within the VSA cold spot  compared with those in clusters
and intercluster regions. Conclusions are drawn in Section~5.

\section{The Corona Borealis Supercluster of galaxies}

One of the best  examples of known superclusters in the northern sky is the
CrB-SC. Abell  (1958) first noted the presence of a concentration of clusters of
galaxies in the Corona Borealis region, and included it in  his catalogue of
second order clusters.  The first dynamical study of the CrB-SC was carried out by
Postman et al.\ (1988) through the  analysis of a sample of 1,555 galaxies in the
vicinity of Abell  clusters. They concluded that the mass of  each cluster in the
core of the CrB-SC lies in the range $1.5-8.9~\times~10^{14}~M_{\odot}$, while the
total mass of the supercluster is   $\approx 8.2~\times~10^{15}~M_{\odot}$, which
is probably enough to bind the system. They proved also that the dynamical
timescales are comparable with the Hubble time, making it unlikely that the system
could be virialized. Subsequent work  by Small et al.\ (1998)  quoted a
value for the mass of CrB-SC of $3-8~\times~10^{16}\ h^{-1}~M_{\odot}$. They
remarked that almost one third of the galaxies in the region are not associated with
any Abell  cluster and noted the great contribution to the projected surface
density of galaxies of the background cluster Abell  2069 and its surrounding
galaxies. This cluster is located at a redshift $z~\approx 0.11$, suggesting the existence of the
so-called Abell  2069 supercluster. The number of clusters belonging to CrB-SC
ranges from six to eight according to Postman et al.\ (1988)  and  Small et al.\
(1997)  respectively. We will consider here that CrB-SC includes
eight clusters (Einasto et al.\ 2001) distributed  around the position
RA=$15^{h}25^{m}16.2^{s}$, Dec.=$+29^{\circ} 31'30''$, at a redshift $z\sim0.07$. 
Six clusters  (Abell  2061, 2065, 2067, 2079, 2089 and 2092) are located in the
core of the supercluster, in  a $\sim 3^{\circ} \times~3^{\circ}$ region, while
there are another two  (Abell  2019 and 2124) at an angular distance of $\sim
2.5^{\circ}$ from the core. Einasto et al.\ (2001) pointed out out  Abell  2061 and
2065 as  the only X-ray emitting clusters. There are four other Abell  clusters (Abell  2056,
2005, 2022 and 2122) in this  region at redshifts  $z\sim0.07$, two (Abell 
2069 and 2083) at $z\sim0.11$ and another two, Abell  2059  and 2073, in the background at redshifts
0.13 and 0.17 respectively. Other clusters have been detected using the Digital Sky
Survey  (Gal et al.\ 2003; Lopes et al. 2004), the Hamburg/RASS catalogue (Bade et
al. 1998), and Sloan Digital Sky Survey (SDSS)  (Koester et al. 2007), all of them
with just photometric estimates of redshifts. Throughout this paper we adopt for CrB-SC a centre
of RA=$15^h 25^m 16.2^s$, Dec.=$+29^\circ 31'30''$ and a radius of 2.5 degrees, which
corresponds  to $\sim 12.5$ Mpc at the distance of that supercluster.

\section{SDSS data}

\begin{figure}
\includegraphics[width=\columnwidth]{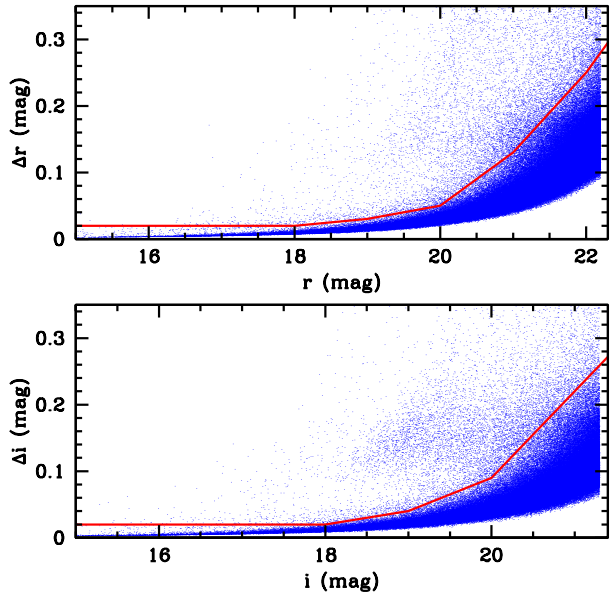}
\caption{Errors of galaxies in SDSS as a function of their magnitude in $r$ and $i$. 
Only objects with errors below the solid lines have been used for the analysis presented 
in this
paper. }
\label{errors}
\end{figure}

\begin{figure}
\includegraphics[width=\columnwidth]{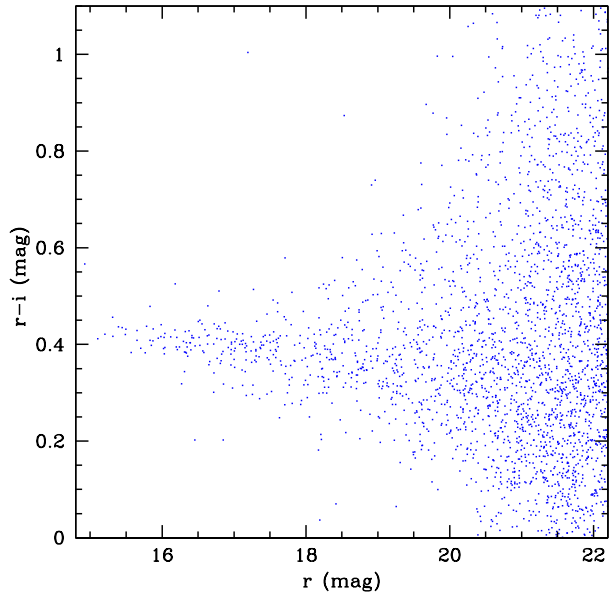}
\caption{Colour--magnitude diagram for SDSS galaxies at distances $\le 15$ arcmin from the centre
of the cluster Abell 2065. }
\label{a2065}
\end{figure}

Using the \emph {SDSS-DR6 Catalogue Archive
Server}\footnote{http://www.sdss.org/dr6/access/index.html\#CAS} by SQL queries, we got the
catalogues of galaxies selected from the {\it  SDSS Galaxy}  file. The SDSS photometric
survey covers the entire region of the  supercluster, apart from two narrow strips (see
below). We downloaded this catalogue in the region $220^\circ \le RA \le 240^\circ$ and
$22^\circ \le$Dec.$\le 37^\circ$  square degrees, and then centred in the VSA cold
spot. There were 5,692,792 objects classified as galaxies in this area. The  extinction
in an area of $\sim 1$ square degrees around the VSA cold spot is small (0.06--0.08 mag in
the $r$ filter) producing reddenings $0.01\le (r-i)-(r-i)_0\le 0.04$. In any case, the
analysis presented in this paper uses the unreddened $r$ and $i$ magnitudes. We limit
our analysis to galaxies with $r\le 22$ mag, which roughly corresponds to the completeness
limits for pointlike objects  in the SDSS catalogue (although some of the analysis use
subsamples with additional cuts in magnitude). At the redshift ($z=0.07$) of the
CrB-SC, $r=22.2$ mag corresponds to $M_r=-15.3$ mag, and then our selection includes
the full range of giants and $\sim 2$ mag within the domain of dwarf galaxies in the
Corona supercluster. The most distant galaxies within the catalogue are expected to be
at $z\sim 0.7$,  where the faintest galaxies  would have $M_r=-21$ mag.
With $r\le 22$ mag the catalogue contains  3,282,036  galaxies. We
imposed an additional constraint according to the photometric errors. Figure~\ref{errors}
presents the errors as a function of the magnitude in $r$ and $i$  filters. As
expected, for most of the objects there is a tight correlation between magnitude and
error. We selected objects with errors below the solid lines in Fig~\ref{errors} (see
Stoughton et al. 2002 for a discussion of photometric properties of SDSS data).
This constraint removes $\sim 5$ \% of the objects, reducing the final number to
3,110,085 in the $r$ filter. For some of the analyses presented in this paper, we
used a further restriction in colour $r-i$, which is required to be in the range 0.2--0.6 according to
the expectations for galaxies in the red sequence (Gladders \& Yee 2000) of Corona Borealis clusters. This is
illustrated in 
Fig.~\ref{a2065}, which shows a colour--magnitude diagram for galaxies at distances $\le 15$ arcmin from the centre
of the cluster Abell 2065. With
this restriction there are 1,217,025 galaxies left in the sample. In addition, for some of
the analyses presented here we use the full sky SDSS catalogue  with all galaxies  up
to $r=19$ mag in a region of $\sim 10,000$ square degrees.

\section{Analysis}

\subsection{The VSA cold spot}

\begin{figure}
\includegraphics[width=\columnwidth]{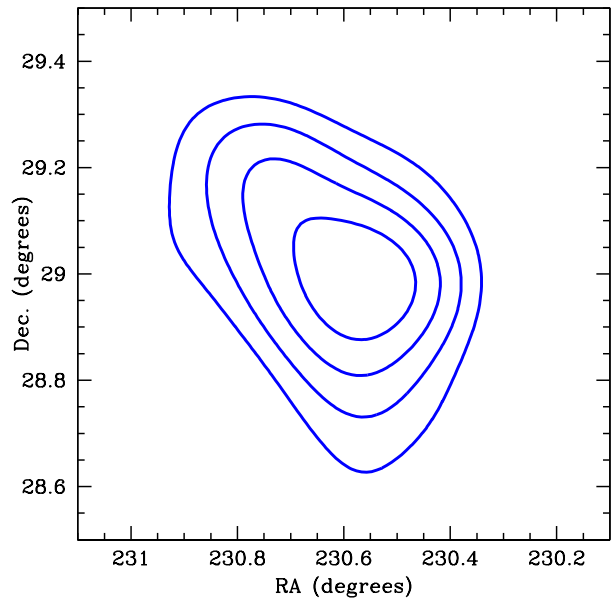}
\caption{Contours of the VSA cold spot (see G08 for details) in the CrB-SC obtained using the superextended
configuration (the effective resolution is 13 arcmin).
The contours are at $-$36.2, $-$54.3, $-$72.4 and $-$90.5   mJy beam$^{-1}$ which correspond to 2, 3, 4 and
5 $\sigma$ levels respectively.}
\label{mancha}
\end{figure}

The VSA observed the CrB-SC with two different setups providing angular resolutions of 11
and 7 arcmin respectively. The cold spot was detected with both configurations. The
position, size and shape of the cold spot have been determined from the uv-tapered map
of the superextended configuration which has an effective resolution of 13 arcmin.  The
cold spot was  the only feature in that map above the $\sim 2\sigma$ level.
Figure~\ref{mancha} presents a contour map with the 2, 3, 4 and 5 $\sigma$ levels of the
cold spot obtained from  the VSA map after source subtraction and Gaussian filtering
(bottom-left of Fig. 1 in G08). The spot is resolved by the VSA and has an
elongated shape with the major axis running along the NE--SW direction. The minimum 
temperature is achieved  at RA=$15^h 22^m 11.47^s$,
Dec.=$29^\circ 00^\prime 02.6^ {\prime \prime}$. The area subtended
by the 2, 3, 4 and 5 $\sigma$ levels are 894, 548, 311 and 120 square arcmin
respectively. The spot is located $\sim 50$ arcmin SW  from the nominal coordinates of
CrB-SC as given by
NED.\footnote{http://nedwww.ipac.caltech.edu/} 

The densities of galaxies within the 2, 3, 4 and 5$\sigma$  (36.2, 54,3, 72.4 and 90.5
mJy beam$^{-1}$ respectively) VSA contours are listed in
Table~\ref{tab1}. We have applied the restriction in errors explained in the previous section and
an $r=22$ mag cut in magnitude. The density of galaxies tends to be slightly higher towards the
inner part of the spot; for instance it increases  $\sim 16$ \% from the 2 to the 5 $\sigma$
contour level. The significance and consequences of this change are investigated in the following
sections.  The small changes ($\leq 0.1$ mags) in the mean values
of  colours $r-i$ are compatible with the dispersion in the distribution of galaxies within
each contour. 

\begin{table}
\caption{Galaxies within the VSA cold spot. The values correspond to  galaxies $15<r<22$ mag within
the 2--5 $\sigma$ contour levels. For each contour level, columns (2--5) show the
 area subtended, 
number , density and mean $r-i$ colour  respectively of galaxies within those contour levels.}
\begin{tabular}{cccccccc}
\hline
$\sigma$-level & area (arcmin) & gal & $\rho$ (gal arcmin$^{-2}$) & $\overline{r-i}$ \\
\hline
  2 & 894 & 2565 & 2.87 & 0.428  \\
  3 & 548 & 1614 & 2.95 & 0.426  \\
  4 & 311 &  978 & 3.14 & 0.430  \\
  5 & 120 &  398 & 3.32 & 0.439  \\
\hline
\end{tabular}
\label{tab1}
\end{table} 

\subsection{The distribution of galaxies within CrB-SC}

\begin{figure}
\begin{center}
\includegraphics[width=8.1cm,height=7.4cm]{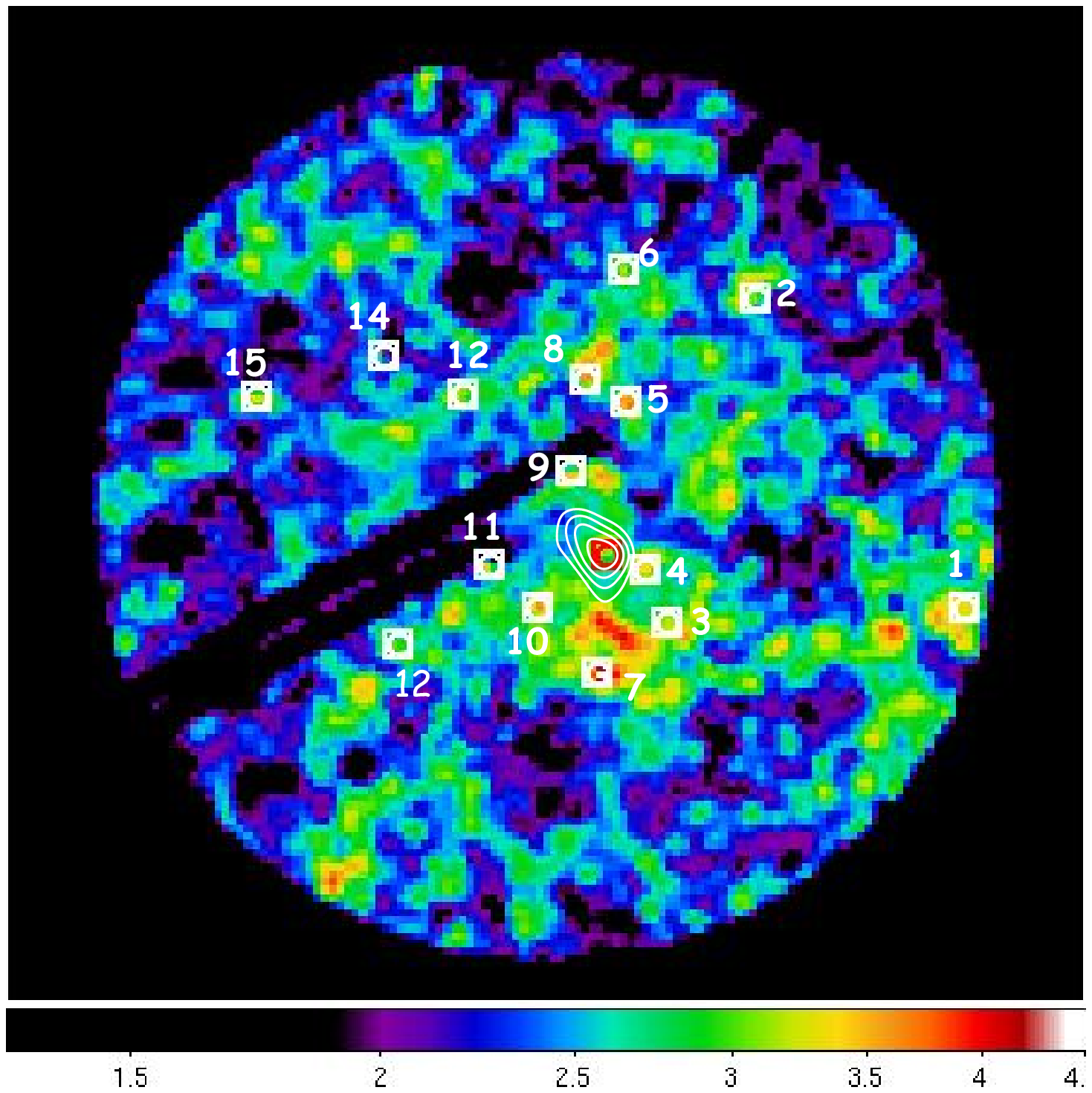}
\includegraphics[width=8.1cm,height=7.4cm]{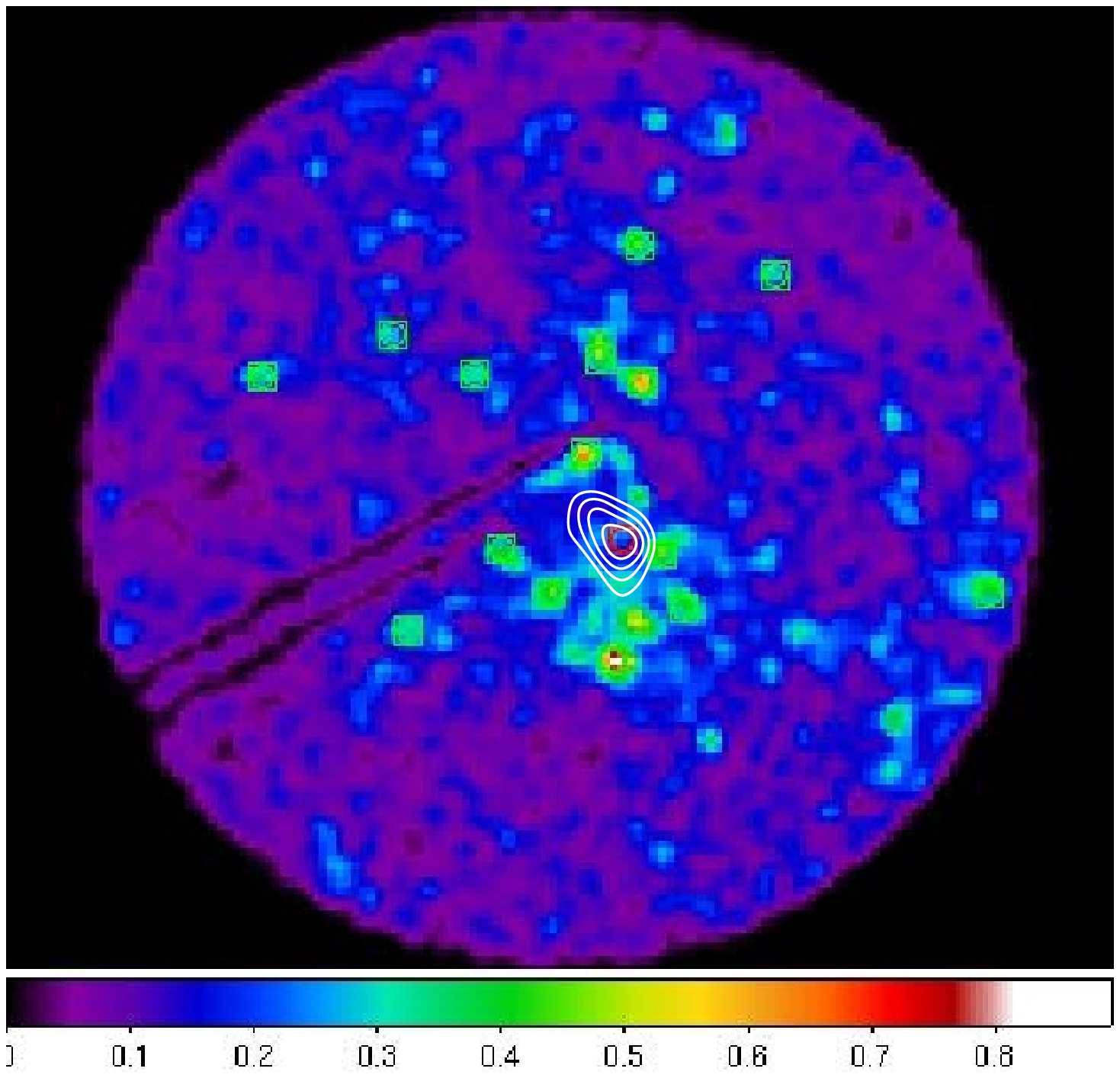}
\caption{Density of galaxies in the CrB-SC region. The maps have been built by counting SDSS 
galaxies in square boxes of $5^\prime \times 5^\prime$ arcmin with the restrictions $15 \le
r\le 22$ ($top$) and $15 \le r\le 19$ ($bottom$). North is up and east is to the left. The maps
have a radius of 5 degrees and are centred in RA=$15^h 25^m 16.2^s$, Dec.=+29$^\circ$ 31' 30''.
The  dark stripes running from the SE to the centre of the maps correspond to regions not
covered by the SDSS catalogues. The white curves correspond to the contours of the VSA cold
spot.   Labels  indicate the positions of the known Abell
clusters  (1. Abell 2022; 2. Abell 2049; 3. Abell 2056; 4. Abell 2059; 5.
Abell 2061; 6. Abell 2062; 7. Abell 2065; 8. Abell 2067; 9. Abell 2069; 10. Abell 2073; 11.
Abell 2079; 12. Abell 2083; 13. Abell 2089; 14. Abell 2092 and 15. Abell 2110).
}
\label{maps1}
\end{center}
\end{figure}

Figure~\ref{maps1} shows the density map of SDSS galaxies in the CrB-SC with $15\le r\le 22$,
and $15\le r\le 19$ respectively.  The maps have been built by counting galaxies within
 $5^\prime \times 5^\prime$ square boxes. The maps have a radius of 5 degrees and are
centred at $RA=15^h25^m16.2^s$, $Dec.=+29^\circ 31'30')$. The dark strips running from
the east edge to the centre of the  maps indicate a region not covered by the SDSS DR6
data.  The highest density regions  in the maps correspond to the known Abell 
clusters. There are no Abell  clusters of galaxies in the area covered by the VSA
cold spot ($\sim 540$ square arcminute below the 3$\sigma$ level). The nearest  cluster
to this spot is Abell  2059  which is projected at 26.2 arcmin (3.6 Mpc at the redshift
$z=0.1305$ of this cluster). The nearest cluster member of CrB-SC is Abell  2056
projected at 58.9 arcmin. This corresponds to  4.7 Mpc ($\sim 3$ Abell  radius) at the
redshift of the supercluster, so there is no appreciable contribution of
galaxies within this cluster at the VSA position.  The highest intercluster density is
found in a region roughly enclosed by a circle with a radius $\sim 1$ degree and
centred  $\sim 40$ arcmin south of the VSA cold spot. The limits of that region are
roughly delineated by the Abell  clusters 2056, 2065, 2059 and 2073. A filament centred
at $RA\sim 15^h22^m$,  $Dec.=28^\circ 10'$ and running from Abell  2065 in the NE--SW
direction  strip with a projected length of $\sim 35$ arcmin (3 Mpc at the redshift of
the CrB-SC) shows the highest density of galaxies not specifically associated with a
cluster. In the following sections we quantify the density and properties of galaxies
within the area subtended by the VSA cold spot as compared with that in cluster and
intercluster regions, and in the control field.  

\subsection{Comparison with control fields}

\begin{figure} 
\includegraphics[width=\columnwidth]{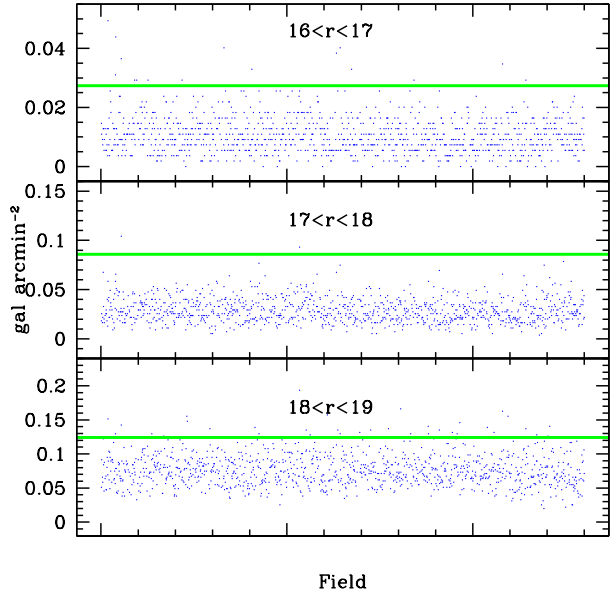}
\caption{Density of galaxies for three ranges in $r$ magnitude within the region subtended
by the VSA cold spot (solid line), as compared with values in 1000 random regions in the
field.} 
\label{compa} 
\end{figure}

We have compared the density of galaxies within the VSA cold spot with those in $1000$
randomly selected regions in the SDSS. For simplicity we computed the density within circles with
a radius of 13.2 arcmin, which gives an  equivalent area to that subtended by the 3$\sigma$
VSA cold spot contour. The fields are distributed across  the whole sky region surveyed
by the SDSS and are far away ($>$30 arcmin) from any Abell cluster.  Figure~\ref{compa} presents the density of galaxies in each of these fields
for three different cuts in $r$ magnitude. The density within the VSA cold spot is indicated by
the solid line. The figure shows that for each cut in $r$ magnitude the great  majority of
fields have densities with values lower than that within the VSA cold spot. The results
are quantified in Table~\ref{tab2}. Only $\leq 1-3$ \% of fields have densities 
comparable to or larger
 than the one in the VSA cold spot. The contrasts in density between VSA and the
field are  $\geq 2$. This overdensity is significant at the 2--4 $\sigma$ level depending
on the cut in $r$ magnitude applied.

\begin{table}
\caption{Density of galaxies within  $1000$ randomly selected regions in the sky as compared with the one
in the VSA cold spot.  Columns are: (1) range in $r$ magnitude analysed; (2--3) density of galaxies 
in the VSA cold spot and mean values in the field; (4) number of fields out of 1000 analysed
in which the density of galaxies is higher than in the region subtended by the VSA cold
spot;  (5) Contrast in density between
VSA and the mean value in the field; (6) Statistical signficance of the VSA overdensity.}
\begin{tabular}{cccccc}
\hline
$r$  &  ${\rho}_{\scriptsize VSA}$ & $\bar{\rho}_{field}$ & $>$VSA & $\Delta \rho/\rho$
&$\sigma$ \\
(mag) & (gal arcmin$^{-2}$) & (gal arcmin$^{-2}$) \\
\hline
16-17 & 0.027 & $ 0.011\pm 0.006$ & 16 & 2.5 & 2.6 \\
17-18 & 0.086 & $ 0.029\pm 0.012$ &  2 & 3.0 & 4.7 \\
18-19 & 0.124 & $ 0.077\pm 0.022$ & 32 & 1.6 & 2.1 \\
\end{tabular}
\label{tab2}
\end{table} 

\subsection{Comparison with CrB-SC intercluster regions}

\begin{figure}
\includegraphics[width=\columnwidth]{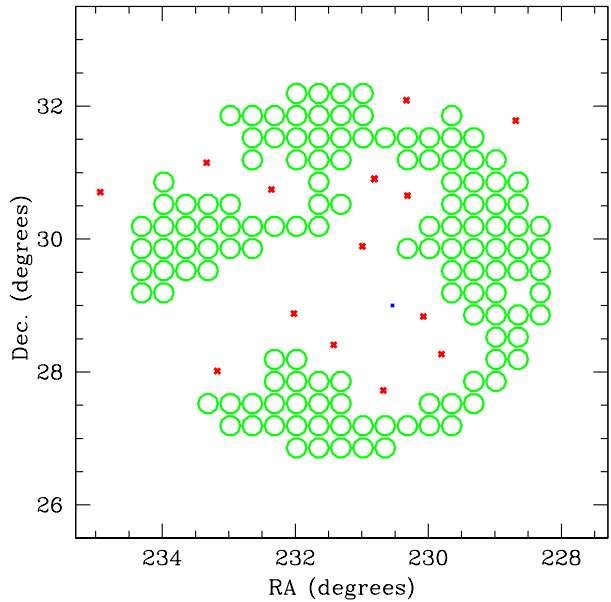}
\caption{Circle with a radius of 3 degrees centred on the CrB-SC centre
RA=$15^h 25^m 16.2^s$, Dec.=+29$^\circ$ 31' 30''. Green circles indicate the position of
the regions used to estimate the intercluster density in CrB-SC. Red crosses indicate the position
of Abell  clusters. The blue square indicates the position of the VSA cold spot.}
\label{pos}
\end{figure}

To estimate the density of galaxies in CrB-SC intercluster regions we built a grid with circles
of 13.2 arcmin (the effective radius of the 3$\sigma$ VSA cold spot contour level) separated 20
arcmin in RA  and Dec., and uniformly covering  the CrB-SC up to 3 degrees from the centre of
the supercluster. We then excluded  those circles  at less than 30 arcmin from the centre of
the VSA cold spot, or at  less than 15 arcmin from an Abell  cluster, or only partially covered
by the SDSS. These restrictions implicitly exclude the inner core of the CrB-SC where the
intercluster density is the highest within the maps presented in Figure~\ref{maps1}. The
position of the 133 remaining regions are shown in Figure~\ref{pos}. The total intercluster
area used for the comparison is then 20.2 square degrees (70\% of the area subtended by a
circle with a radius of 3 degrees centred at the CrB-SC). The results are summarized in
Table~\ref{tab3}, which presents the mean number of galaxies and $r-i$  colour in the
intercluster regions and in the VSA cold spot. The region subtended by the VSA cold spot is
overdense by factors $\sim 2$ for any of the considered cuts in magnitudes. These factors are
in the range3--4$\sigma$ level of significance and  are very robust with respect to magnitudes
or colour restrictions. The mean value of $r-i$ colours in intercluster regions and within the
VSA cold spot are compatible when dispersion from field to field and Poissonian noise are taken
into account. The overdensity of galaxies found within the VSA cold spot appears to be
consistent with  previous findings of potential clusters of galaxies (Gal et al. 2003; 
Koester et al. 2007) at  a distance of 8-9 arcmin from the center of the spot.

\begin{table*}
\caption{Number of galaxies and mean colour $r-i$ within the area subtended by the VSA
cold spot (cols 1--2 and 5--6), and mean values in the intercluster CrB-SC regions (cols
3--4 and 7--8). These values have been
computed without restrictions in colours (cols 1--4), and with a restriction $0.2\le r-i
\le 0.6$ (cols 5--8).}
\begin{tabular}{ccccccccc}
\hline
& (1) & (2) & (3) & (4) & (5) & (6) & (7)& (8) \\
\hline
Range in mag  & gals & $\overline{r-i}$ & gals & $\overline{r-i}$ & gals & $\overline{r-i}$ & gals & $\overline{r-i}$  \\
\hline
$18<r<19$  &   74 &  0.380 & $ 32.05\pm  12.32$  & $0.412\pm 0.060$ &  67 & 0.389 & $ 28.74\pm  11.65$ & $0.417\pm 0.024$ \\
$19<r<20$  &  151 &  0.424 & $ 86.91\pm  22.78$  & $0.420\pm 0.040$ & 129 & 0.391 & $ 70.03\pm  20.09$ & $0.411\pm 0.017$ \\
$20<r<21$  &  380 &  0.435 & $231.77\pm  49.84$  & $0.428\pm 0.031$ & 243 & 0.385 & $145.74\pm  34.38$ & $0.396\pm 0.012$ \\
$21<r<22$  &  897 &  0.429 & $582.38\pm 109.87$  & $0.404\pm 0.033$ & 417 & 0.388 & $262.59\pm  50.65$ & $0.392\pm 0.009$ \\ 
\hline
\end{tabular}
\label{tab3}
\end{table*}

\subsection{Comparison with CrB-SC clusters}

\begin{table*}
\caption{Density of galaxies in several fields in Corona Borealis Supercluster: The regions
are centred at the  VSA cold spot, the Abell  clusters at distances up to 3 degrees 
from that spot (the clusters Abell  2069, 2079  and 2092 have not considered here as
they are not totally covered by the SDSS photometric  catalogue), and five control fields where there are no Abell  clusters. The galaxies
are selected  with photometric errors  as explained in Section~2. It is
shown the density within circles of 5, 10, 15 and 20 arcmin respectively, for cuts in
magnitude $r<19$ (cols 4--7), and $r<22$ (cols 8--11).} 

{\tiny 
\begin{tabular}{lccccccccccccc}
\hline
Field& RA (J2000) & Dec. (J2000) & Redshift & & & & $\rho$ (gal arcmin$^{-2}$) & \\
& & & & & $15 \le r \le 19 mag $    & & &  & $15 \le r \le 22$ mag   \\
\hline
 & (hh mm ss.ss)& (dd mm ss.s)& & $R=5'$ &$R=10'$ & $R=15'$ & $R=20'$ & $R=5'$ &$R=10'$ & $R=15'$ &
 $R=20'$ \\
\hline
		   	 								   
VSA cold spot&  15 22 11.5 & +29 00 06 &	    &	0.253  &   0.255 &    0.215 &	 0.163 &  3.214&     2.888  &	2.606  &   2.335\\
Abell  2073   &  15 25 41.5 & +28 24 32 & 0.171700   &	0.594  &   0.335 &    0.260 &	 0.200 &  3.727&     3.050  &	2.593  &   2.652\\
Abell  2059   &  15 20 17.6 & +28 50 13 & 0.130500   &	0.560  &   0.331 &    0.214 &	 0.231 &  3.298&     3.058  &	2.862  &   2.543\\
Abell  2061   &  15 21 15.3 & +30 39 17 & 0.078400   &	0.904  &   0.450 &    0.190 &	 0.150 &  4.598&     2.986  &	2.392  &   2.266\\
Abell  2067   &  15 23 14.8 & +30 54 23 & 0.073858   &	0.574  &   0.383 &    0.186 &	 0.194 &  3.811&     3.062  &	2.644  &   2.542\\
Abell  2083   &  15 29 26.3 & +30 44 45 & 0.114200   &	0.361  &   0.309 &    0.162 &	 0.140 &  3.232&     2.869  &	2.582  &   2.347\\
Abell  2056   &  15 19 12.3 & +28 16 10 & 0.084600   &	0.429  &   0.382 &    0.245 &	 0.200 &  3.066&     3.120  &	2.891  &   2.807\\
Abell  2065   &  15 22 42.6 & +27 43 21 & 0.072600   &	1.021  &   0.541 &    0.303 &	 0.255 &  4.449&     3.771  &	2.878  &   2.991\\
Abell  2089   &  15 32 41.3 & +28 00 56 & 0.073130   &	0.400  &   0.292 &    0.161 &	 0.145 &  2.940&     2.763  &	2.536  &   2.309\\	    
\hline        
\hline
\end{tabular}
}
\label{tab4}
\end{table*} 

Table~\ref{tab4} shows the density of galaxies in the VSA cold spot and in the Abell 
clusters located up to  3 degrees from the centre of the CrB-SC. The clusters Abell  2069,
2079 and 2092 have been removed from that list because they are not completely covered
by the SDSS. The table shows the density within radius of 5, 10, 15 and 20 arcmin 
respectively for $15 \le r\le 19$ and $15 \le r\le 22$ mags respectively.  The
density for Abell  clusters within $R\le 10$ arcmin for the deeper cut in magnitude, is
in the range 2.763--3.771  arcmin$^{-2}$. The cluster with the higher density of
galaxies is Abell  2065 as expected from the fact that it is one of the two cluster in
the field with X-ray emission in ROSAT. 

\begin{figure}
\includegraphics[width=\columnwidth]{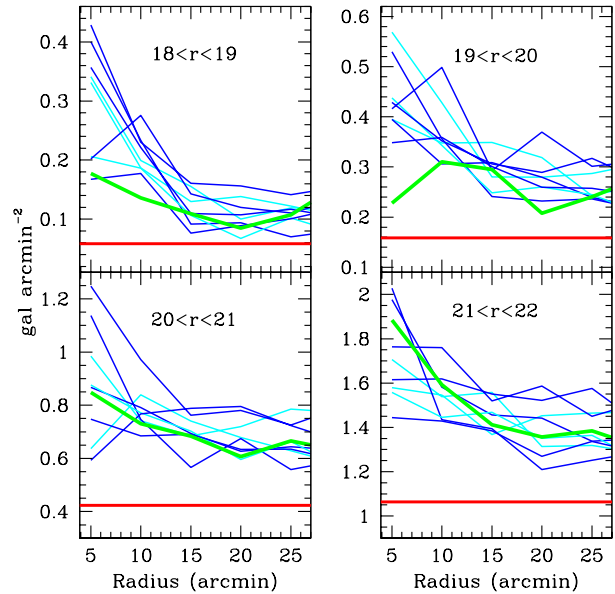}
\caption{Density of galaxies for several cuts in $r$ magnitude as a function of radius for 
the VSA cold spot (green line), Abell  clusters
of galaxies  (blue lines), and the mean value in intercluster regions (red horizontal 
line).}
\label{radial}
\end{figure}

Figure~\ref{radial}
presents the radial dependence of such fields for several ranges in magnitude. The scale of the overdensity associated with the VSA cold spot
region is 10--15 arcmin, which is similar to the typical radius of clusters at the
redshift of CrB-SC. The excess found in the  $18<r<19$ mag slice could be interpreted
as the existence of a relatively low mass (or relatively dark) cluster at the position of the VSA
cold spot and
at the redshift of CrB-SC. However, the absence of overdensity in the  $19<r<20$ slice
suggests that this concentration of galaxies  cannot be  responsible for the overdensity
 at fainter magnitudes.

\begin{figure}
\begin{center}
\includegraphics[width=13cm,height=10cm]{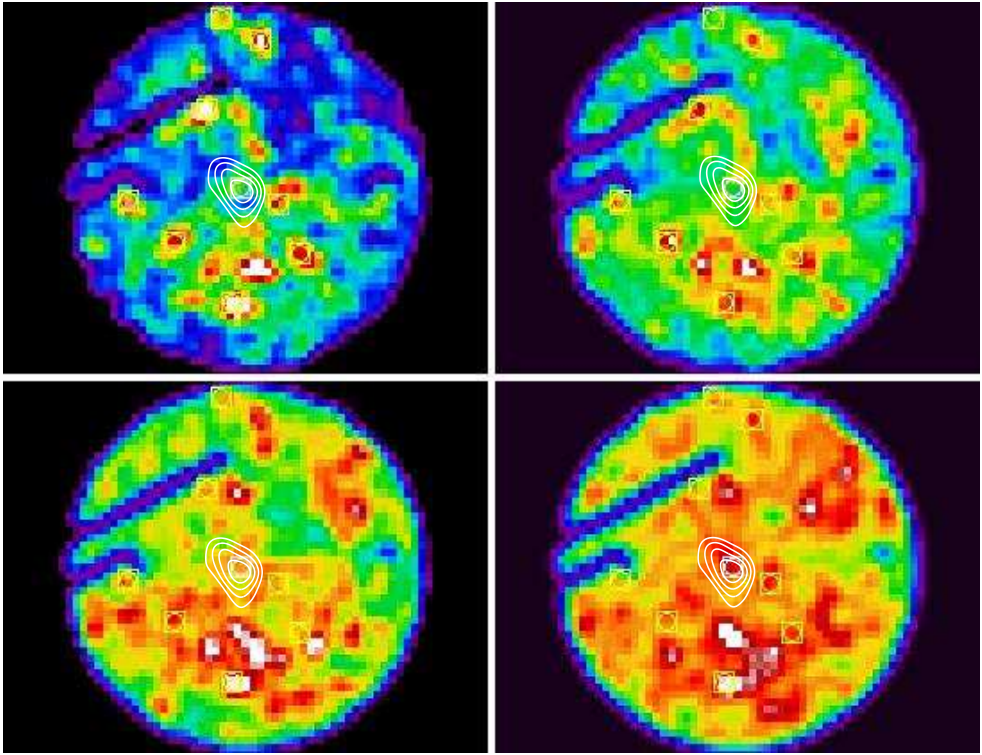}
\caption{Density of galaxies in a region with a radius of 2.5 degrees centred in the
position of the VSA cold spot (see Fig.~\ref{maps1} for details). Each panel shows a
different slice in $r$ magnitude: top-left $18<r<19$;  top-right $19<r<20$; bottom-left
$20<r<21$,  and bottom-right $21<r<22$.}
\label{maps2}
\end{center}
\end{figure}

\begin{figure}
\begin{center}
\includegraphics[width=15cm,height=10cm]{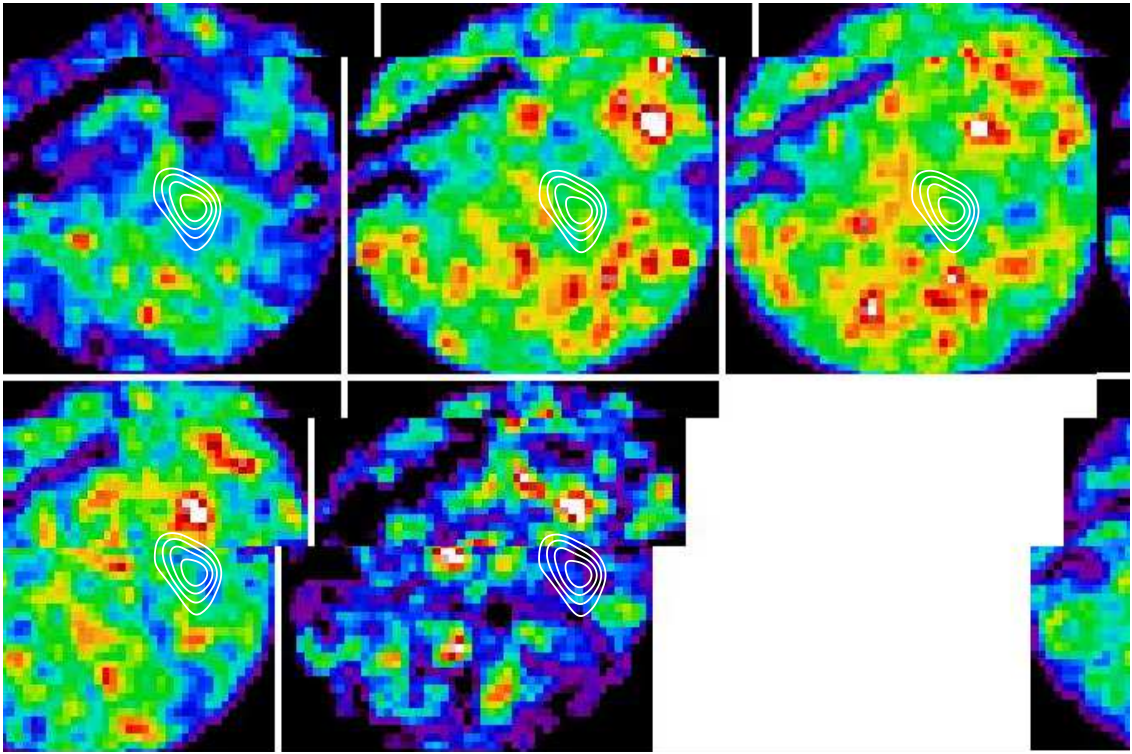}
\caption{Density of galaxies in a region with a radius of 2.5 degrees centred in the
position of the VSA cold spot (see Fig.~\ref{maps1} for details). Each panel shows a
different slice in $r-i$ colour (from top-left to bottom-middle): $0.3<r-i<0.5$,
$0.5<r-i<0.7$, $0.7<r-i<0.9$, $0.9<r-i<1.1$ and $1.1<r-i<1.3$ respectively.}
\label{maps3}
\end{center}
\end{figure}

Figure~\ref{maps2} shows maps of the density of galaxies in a circle with a radius of 2
degrees centred at the position of the VSA cold spot. Each panel shows a slice which
corresponds to a selection of galaxies in $r$ magnitude. As shown in
Fig.~\ref{maps1}, the spot is immersed within the inner core of the supercluster where
the density is higher. Abell  clusters are the dominant structures in the maps which
show the brighter slice in magnitude. As we move to fainter cuts in magnitude, the
presence of those clusters is less obvious as more and more galaxies from the background
are contributing to the maps. In the first three panels the region subtended by the VSA
cold spot is within a filament running from SE to NW which is slightly overdense with
respect to the near  environment. However, the map  corresponding to the 
$21<r<22$ mag slice (bottom right) shows a clear overdensity with a radius $\sim 15$ arcmin 
located at the position of the VSA cold spot. The absence of that overdensity in the 
$19<r<20$ slice could indicate the presence of a cluster in CrB-SC with a lack of
relatively bright galaxies, or alternatively that the concentration of  galaxies is in
the background of the
respect to CrB-SC. Figure~\ref{maps3} shows the density of galaxies up to  $r=22$ mag for
different $r-i$ colour slices. In the redder map (bottom-middle panel) a clear
overdensity emerges at the position of the VSA cold spot. This corresponds basically to
the structure seen in the $21<r<22$ mag in Figure~\ref{maps2}.

\section{Conclusions}  

In this paper we have presented a study of the spatial distribution of galaxies in the
CrB-SC, where a cold spot at 33 GHz was found by the  VSA instrument working with two
different  configurations. We have used the SDSS data to analyse the density, and
photometric properties of galaxies within the area covered by this spot and compared
them with those of galaxies in Abell  clusters, intercluster regions and in the field.
Our main findings are:

\begin{itemize}

\item The VSA cold spot is located in the northern part (40 arcmin from the centre) of a
region of $\sim 1$ degree delineated by the clusters Abell  2056, 2065, 2059 and 2073, 
where the density of galaxies is a factor $\sim 2$ higher with respect to other
intercluster regions. This gives an excess  of $\sim 8\times 10^3$ galaxies up to $r=22$ 
mag  with respect to typical values in the field.

\item The region subtended by the VSA cold spot has an overpopulation of galaxies as
compared with mean values in random selected areas of the sky. Counting galaxies up to
$r=19$ mag in $\sim 1000$ random selected regions in the sky, we showed that the
density of galaxies within the spot is $\sim$ 1.6--3.0 higher than the mean value in such
control fields with little dependence on the restrictions applied in magnitudes and/or
colours. Only in $\sim$ 1--3\% of such random regions the density of galaxies is larger
than in the VSA cold spot. 

\item No Abell  cluster of galaxies is found in the area subtended by the VSA cold
spot ($\sim 540$ square arcminute below the 3$\sigma$ level). The nearest Abell  cluster
is Abell  2059, which has a redshift of 0.1305 and is projected at 26.2 arcmin (3.6 Mpc at the
redshift $z=0.1305$). The nearest cluster member of CrB-SC is Abell 
2056 projected at 58.9 arcmin (4.7 Mpc at the redshift of the supercluster). We
conclude  that the contribution of any Abell  cluster to the population of galaxies
within the area subtended by the VSA cold spot is probably very small.

\item The area subtended by the VSA cold spot shows an excess of faint ($21<r<22$) and
red  ($1.1<r-i<1.3$) galaxies, as compared with typical values within CrB-SC intercluster
regions. This overdensity of galaxies shows a radial dependence and extends up to $\sim
15$ arcmin. This could be the signature of a previously unnoticed cluster at high redshift.
Confirmation and delineation of such 
structure will 
require deep imaging and spectroscopic determination of redshifts. 

\end{itemize}

\onecolumn

\twocolumn

\end{document}